\documentstyle{article}

\title{Complementarity in the Bohr-Einstein Photon Box}
\author{Dennis Dieks and Sander Lam\\ History and Foundations of Science \\
Utrecht University, P.O. Box 80.000\\ 3508 TA  Utrecht,
The Netherlands}
\begin{document}
\date{}
\maketitle
\begin{abstract}
The photon box thought experiment can be considered a forerunner
of the EPR-experiment: by performing suitable measurements on the
box it is possible to ``prepare'' the photon, long after it has
escaped, in either of two complementary states. Consistency
requires that the corresponding box measurements be complementary
as well. At first sight it seems, however, that these measurements
can be jointly performed with arbitrary precision: they pertain to
different systems (the center of mass of the box and an internal
clock, respectively). But this is deceptive. As we show by
explicit calculation, although the relevant quantities are
simultaneously measurable, they develop non-vanishing commutators
when calculated back to the time of escape of the photon. This
justifies Bohr's qualitative arguments in a precise way; and it
illustrates how the details of the dynamics conspire to guarantee
the requirements of complementarity. In addition, our calculations
exhibit a ``fine structure'' in  the distribution of the
uncertainties over the complementary quantities: depending on
\textit{when} the box measurement is performed, the resulting
quantum description of the photon differs. This brings us close to
the argumentation of the later EPR thought experiment.
\end{abstract}

\section{Introduction}
The 1930 Solvay conference was the scene of the famous
weighing-of-energy debate between Einstein and Bohr. According to
Bohr's report \cite{bohr}, the debate revolved around the validity
of the time-energy uncertainty relation. As Bohr tells us,
Einstein had devised an ingenious thought experiment (involving a
``photon box'') with which he wanted to demonstrate that an
individual photon can have both a sharply defined energy and a
precisely predictable time of arrival at a detector. If
successful, this would mean the demise of the time-uncertainty
relation. Einstein himself later maintained that Bohr had
misunderstood his intentions: that it was not the validity of the
uncertainty relation, but rather the unpalatable implications of
complementarity in the case of correlated distant systems that he
was targeting. Considered either way, the thought experiment
furnishes a remarkable illustration of quantum mechanical
complementarity.

The idea of the experiment is to start with a box, filled with
radiation, hanging stationary in the gravitational field (after a
preliminary balancing procedure). In this situation the total
energy of the box and its contents has a well-defined
value\footnote{Since the box system has a very large mass, the
spread in energy can be vanishingly small even though the spreads
in position and momentum of its center of mass do not vanish.}.
Inside of the box there is a clock that opens a small shutter when
its hands reach a fixed position. This shutter remains open for a
very brief time interval, during which one photon escapes. After
the photon's escape (possibly at a much later time), the box is
weighed, so that its mass---and therefore its energy---can be
determined. Comparison with the initial situation gives us the
energy of the escaped photon. In addition we can read off the
internal clock, and this will tell us how much time has elapsed
since the opening of the shutter.

The weighing is performed by looking at how the center of mass of
the box has moved under the influence of gravity since the
photon's escape\footnote{In fact, Bohr suggests to bring the box
back to its zero position by attaching suitable loads. This is
just another way of studying the center of mass quantities.
Therefore, although our calculations are based on a slightly
different weighing procedure, the underlying principle applies
equally to the procedure that was analysed by Bohr.}.
Since the dynamical quantities of the center of
mass commute with the variables of the internal clock, it is
possible to perform a measurement in which both the position of
the hands of the clock \textit{and} the position (or momentum) of
the center of mass of the box are sharply determined. It therefore
appears clear that both the energy and the time of escape of the
photon can be determined with arbitrary precision. That, however,
would imply a violation of the time-energy uncertainty principle
applied to the photon---quantum mechanics must in some way forbid
this joint precise determination. Consistency requires that the
mass of the box and the opening time of the shutter, as determined
from the weighing-plus-reading-off-the-clock measurement, be
complementary quantities. To show exactly how quantum mechanics
makes this happen is the main purpose of this note.

Bohr writes that some time after the original debate
\begin{quote}
``[Ehrenfest] told me that Einstein was far from satisfied and with his
usual acuteness had discerned
new aspects of the situation which strengthened his critical attitude.
In fact, by further examining the possibilities for the application of a
balance arrangement,
Einstein had perceived alternative procedures which, even if they did not
allow the use he originally intended,
might seem to enhance the paradoxes beyond the possibilities of logical
solution.
Thus, Einstein had pointed out that, after a preliminary weighing of the
box with the clock
and the subsequent escape of the photon, one was still left with the choice of
either repeating the weighing or opening the box and comparing the reading
of the clock with the standard time scale.
Consequently, we are at this stage still free to choose whether we want to
draw conclusions
either about the energy of the photon or about the moment when it left the
box.''
(\cite{bohr}, pp.\ 228-229).
\end{quote}

Thus, according to this account Einstein later shifted towards
EPR-like considerations: apparently having been convinced at the
Solvay conference by Bohr's arguments, Einstein now started
pointing out that even if the photon is already far away, we can
still decide to perform one or the other of a pair of
complementary measurements and thus ``prepare'' the photon in
either a state with a well-defined energy or in a state peaked in
time. This is highly remarkable, since we would expect the distant
photon to be a system on its own, with properties that cannot
depend on what happens outside of its lightcone. As already
mentioned, it may be that Bohr misinterpreted the logic of the
sequence of events here: perhaps Einstein already accepted the
validity of the uncertainty relations in 1930 and meant his photon
box experiment from the start as a kind of delayed choice
experiment. This is in fact what is suggested by looking at the
precise text of Ehrenfest's message to Bohr. Ehrenfest visited
Einstein in 1931 and found him to be very outspoken about the
purpose of the photon box. In his letter to Bohr of 9 July 1931,
Ehrenfest reported: ``He said to me that, for a long time already,
he absolutely no longer doubted the uncertainty relations, and
that he thus, e.g., had BY NO MEANS invented the `weighable
light-flash box' `contra uncertainty relation', but for a totally
different purpose'' \cite{how}.

Regardless of whether Bohr's or Einstein's account is closer to
the truth, it is an important point that the box-plus-photon
system furnishes an example of simultaneously existing strict
correlations between complementary quantities. Einstein's thought
experiment can be considered a forerunner of the EPR-experiment.
These general features of the experiment we shall address first
(see also \cite{dieks} for references to earlier discussions of
the photon box).

\section{Global Analysis of the Experiment}

An essential element of the photon-box experiment is the
correlation that exists between box and photon quantities after
the photon has escaped. The relevant photon quantities are its
energy $E_{ph}$ and its time of arrival at a given detector,
$T_{arr}$. These are correlated with the box energy $E$ and the
position of the hands of the clock in the box, $q_{cl}$,
respectively. The two energies are correlated as a consequence of
total energy conservation; and since the clock allowed the photon
to escape when its hands were at a fixed predetermined position,
$q_{cl}=0$, say, after which the photon travelled with the fixed
speed $c$, $q_{cl}$ and $T_{arr}$ are also correlated.

The two photon quantities, energy and time of arrival, are
complementary according to quantum mechanics. Indeed, an energy
eigenstate is a plane wave, which obviously does not have a
well-defined time of arrival at a given point; conversely a very
narrow wave packet contains very many different frequencies, and
therefore is a superposition of different energy states.
Paradoxically, it nevertheless seems that these two photon
quantities can simultaneously be determined, with arbitrary
precision, via judicious measurements on the box.

But if quantum mechanics is consistent, any uncertainty relation
valid for the photon quantities must obviously have its
counterpart in an uncertainty relation for the correlated box
quantities---it should be impossible to beat the uncertainty
relation for photon quantities by measuring correlated box
quantities. We should therefore expect that the box cannot possess
both a definite value of its energy and a definite time at which
the shutter opened. In view of the correlations between the box
and the photon, the total state of these two systems has to be
entangled, such that neither the energies nor the relevant time
quantities of both box and photon are sharply defined, whereas
their correlation \textit{is} a sharply defined quantity
\textit{of the total system}\footnote{This in turn means that the
box and the photon will not have their own pure states: both must
be described by \textit{mixed} states. These mixed states can be
written as mixtures of well-defined mass states, or alternatively
as mixtures of states peaked in time. As long as we discuss the
two systems separately, we may think of these mixtures as
classical ``ignorance mixtures'' without getting into
contradictions.}.

According to relativity, all energy possesses mass and is acted
upon by gravity. The experiment starts with the box in a
stationary position after a preliminary weighing, but after the
escape of the photon the box experiences a net force, which
depends on the mass of the escaped photon, $m$. As a consequence,
the box starts moving and both the position $q$ and the momentum
$p$ of its center of mass get correlated to $m$. A measurement of
either $p$ or $q$ will therefore provide information about $m$. On
the other hand, reading off the time indicated by the clock,
$q_{cl}$, will yield information about how much time has passed
since the shutter was opened. Both $q_{cl}$ and \textit{either}
$p$ \textit{or} $q$ can be measured and can jointly have sharp
values, since these quantities pertain to different systems---the
internal clock and the center of mass of the box,
respectively---and commute.

We have already argued that the escape time calculated from
$q_{cl}$, and on the other hand the box energy as computed from a
measurement of either $p$ or $q$, must be complementary
quantities. This should be reflected by a non-vanishing value of
the commutator of the operators representing these quantities.
That this is indeed so, and that this leads to exactly the right
uncertainty relations, is what we shall show now.

\section{Complementarity of Mass and Escape Time}

We start our considerations with the joint measurement of either
$p$ or $q$ of the center of mass of the box, together with the
clock variable $q_{cl}$. This measurement takes place when the
photon is already well on its way. The chosen center of mass
quantity and the clock variable may both be measured with
arbitrary precision: the corresponding operators commute. However,
in order to be able to say something about the photon, we have to
\textit{calculate back} to values of the relevant quantities at
the time the shutter opened and the photon escaped. To make this
stand out in the calculations below, we choose the origin of time,
$t=0$, at the moment of the final measurement and choose the
positive time direction \textit{backwards}, i.e.\ \textit{going
into the direction of the earlier photon emission event}. We shall
show that the state of the box corresponding to the measurement
result (this state can be thought of as resulting from application
of the projection postulate, or ``collapse of the wavefunction''
to the pre-measurement box state), when followed back in time to
the instant of the photon emission, exhibits quantum spreads that
exactly lead to the expected---and required---uncertainties in
$E_{ph}$ and $T_{arr}$.

Taking into account that during the motion of the box the clock
finds itself at different heights in the gravitational field,
depending on the position of the center of mass, $q$, we have the
following expression for the clock variable at time $t$
\textit{before} the final measurement ($t=0$ corresponds to the
final measurement and $t$ is counted backwards):
\begin{equation}
    q_{cl}(t) =  \int_{0}^t d\tau (1 -
\frac{g q(\tau)}{c^2}).
    \label{clock}
\end{equation}
Bohr took recourse to general relativity to justify the use of
this formula, but since then it has be shown \cite{unruh} that
Eq.\ (\ref{clock}) follows from just the assumption that energy
has mass and can be weighed (which is exactly what Einstein needed
to posit in order to make the thought experiment work in the first
place). This is an important point that makes the analysis of the
experiment self-contained: it would not be satisfactory if only by
invoking general relativity the consistency of quantum mechanics
could be demonstrated.

From (\ref{clock}) we see that
\begin{equation}
    \dot{q}_{cl}(t) =  1-\frac{g}{c^2}q(t).
    \label{clock1}
\end{equation}

For the Heisenberg equations of motion of the vertical position
and momentum of the box we have, with the same convention about
$t$:
\begin{equation}\label{motion}
\dot{q}(t) = p/M\;\;\;\;,\;\;\;\;\dot{p}(t) = -mg -V^{\prime}(q),
\end{equation} where $V(q)$ is the potential in which the box
finds itself (in Bohr's description of the experiment the box is
suspended from a spring, which would correspond to $V(q)=
1/2kq^2$, with $k$ the spring constant); the prime indicates
differentiation with respect to vertical position.

The commutators $[p,q_{cl}]$ and $[q,q_{cl}]$ both vanish at
$t=0$; this means that in the measurement sharp values can be
assigned to both $q_{cl}$ and either $p$ or $q$. But in the
Heisenberg picture the commutators change in time and do not
remain null. For the change of the commutator of $p$ and $q_{cl}$
we can write down the following differential equation:
\begin{eqnarray} \label{comm1}
\frac{d}{dt}[p,q_{cl}]=[\dot{p},q_{cl}]+
[p,\dot{q}_{cl}]=\frac{g}{c^2}i\hbar\ - [V^{\prime}(q),q_{cl}],
\end{eqnarray} where (\ref{clock1}) and (\ref{motion}) have been used.
Similarly, we find for the commutator between $q$ and $q_{cl}$:
\begin{eqnarray} \label{comm2}
\frac{d}{dt}[q,q_{cl}]=[\dot{q},q_{cl}]+
[q,\dot{q}_{cl}]=[\dot{q},q_{cl}]=\frac{1}{M}[p,q_{cl}].
\end{eqnarray}

It follows that in the simplest situation, in which $V^{\prime}$ vanishes
and the box only experiences the force of gravity, we have
\begin{equation}\label{c2}
[p,q_{cl}]=\frac{g}{c^2}i\hbar t
\end{equation} and
\begin{equation}\label{c3}
[q,q_{cl}]=\frac{g}{2Mc^2}i\hbar t^2.
\end{equation}

So the center of mass coordinates develop (recall: backwards in
time!) non-vanishing commutators with the clock variable $q_{cl}$.
Therefore $q_{cl}$ cannot possess a sharp value together with
either $p$ or $q$ at the time of the photon emission!

As a consequence of Eq.\ (\ref{c2}) we have the following
uncertainty relation\footnote{Analogously to the standard
uncertainty relation $\Delta p.\Delta q \geq \frac{1}{2}\hbar$
that follows from the canonical commutation relation
$[p,q]=-i\hbar$.} between $p$ and $q_{cl}$ at time $t$
\begin{equation}\label{unc1}
  \Delta p.\Delta q_{cl} \geq \frac{tg}{2 c^2}\hbar.
\end{equation}
Since in this case, with $V^{\prime}=0$, it follows from (\ref{motion})
that $p(t)=p(0) -mgt$, we have the following relation between the
uncertainties in $p$ and $m$:
\begin{equation}\label{unc2} \Delta p = gt \Delta m.
\end{equation}

This uncertainty in $m$ must be understood in the following way.
As we have just shown, the final measurement result, which may be
completely sharp, translates back to a state with width $\Delta p$
at the time of the photon emission. This introduces an uncertainty
in the determination of $m$: different values of $m$ could have
led to the same final measurement result because the value of
$p(t)$, as ascertained from the final measurement is uncertain.
The value $\Delta m$ just calculated is the range of $m$-values
that is compatible with the width $\Delta p(t)$ and the
measurement result. This $\Delta m$ equals the uncertainty with
which we can make a prediction about the mass of the photon, on
the basis of the measured value $p(0)$. \footnote{Since the state
of the box is a \textit{mixture} of components with different
$m$-values, we may argue about the situation in a classical way:
different possible values of $m$ lead to uncertainty in the
evolution and consequently to an uncertainty in $p$.}

Equation (\ref{unc1}), together with the fact that the uncertainty
in $q_{cl}$ at $t$ equals the uncertainty $\Delta T$ in the
instant of the opening of the shutter and together with $\Delta E=
c^2 \Delta m$, leads to
\begin{equation}\label{U1}
  \Delta E. \Delta T \geq \frac{1}{2}\hbar.
\end{equation}

Alternatively, we may focus on $q$ instead of $p$ in order to
determine $m$. This would be in line with Bohr's account in which
the center of mass of the box is rigidly connected to a pointer
that moves along a scale. In this case we find from
(\ref{motion}): $\Delta q = \frac{gt^2}{2 M} \Delta m$, which
together with (\ref{c3}) leads to
\begin{equation}\label{U2}
  \Delta E. \Delta T = c^2 \Delta m. \Delta q_{cl} = \frac{2 c^2 M}{gt^2}.
\Delta q. \Delta q_{cl} \geq \frac{1}{2}\hbar.
\end{equation}

So regardless of whether we measure $p$ or $q$, we shall not be
able to predict the energy and arrival time of the photon with a
smaller latitude than allowed by the time-energy uncertainty
relation.

If $V^{\prime}$ depends on $q$, the calculations become more complicated.
Let us have a look at the case suggested in Bohr's account, in
which the box is suspended from a spring. We can model this
situation by assuming a harmonic force $-kq$, with $k$ a positive
constant. This leads to the following Heisenberg equations of
motion:
\begin{equation}
\dot{q}(t) = p/M\;\;\;\;,\;\;\;\;\dot{p}(t) = -mg -kq,
\end{equation} and therefore
\begin{equation}
\ddot{q}(t) = -\frac{m}{M}g -\frac{k}{M}q.
\end{equation}
The solutions of these equations are given by:
\begin{equation}
q(t) = \frac{mg}{k}(\cos \omega t - 1) + q(0)\cos \omega t +
\frac{p(0)}{M \omega}\sin \omega t,\label{q}
\end{equation}and
\begin{equation}
p(t) = -\frac{Mm\omega g}{k}(\sin \omega t) - M \omega
q(0)\sin\omega t + p(0)\cos \omega t,\label{p}
\end{equation} in which $\omega^2=k/M$.

For the uncertainties in $m$ connected with $\Delta q$ and $\Delta
p$, respectively, we thus find (for $\omega t << M/m$):
\begin{equation}
\Delta m_{q}(t) = \frac{k}{g(1-\cos \omega t)}\Delta
q(t),\label{mq}
\end{equation}and
\begin{equation}
\Delta m_{p}(t) = \left|\frac{k}{M \omega g \sin\omega
t}\right|\Delta p(t)\label{mp}.
\end{equation}

The commutators $[p,q_{cl}]$ and $[q,q_{cl}]$ in this case, with
$V(q)=\frac{1}{2} k q^2$, satisfy the equations
\begin{eqnarray} \label{comm3}
\frac{d}{dt}[p,q_{cl}]=\frac{g}{c^2}i\hbar\ - k[q,q_{cl}],
\end{eqnarray}
\begin{eqnarray} \label{comm4}
\frac{d}{dt}[q,q_{cl}]=\frac{1}{M}[p,q_{cl}],
\end{eqnarray} so that
\begin{eqnarray} \label{comm5}
\frac{d^2}{dt^2}[p,q_{cl}]=-\frac{k}{M}[p,q_{cl}].
\end{eqnarray}
It follows that
\begin{equation}
[p,q_{cl}]= \frac{g}{c^2} \frac{\sin \omega t}{\omega}i \hbar,
\end{equation}
\begin{equation}
[q,q_{cl}]= \frac{g}{c^2} \frac{1-\cos \omega t}{M \omega^2}i
\hbar.
\end{equation}
This yields the uncertainty relations
\begin{equation}\label{unc3}
  \Delta p.\Delta q_{cl} \geq \left|\frac{g\sin \omega t}{2 \omega
c^2}\right|\hbar,
\end{equation}
\begin{equation}\label{unc4}
  \Delta q.\Delta q_{cl} \geq \frac{g(1-\cos\omega t)}{2 M \omega^2 c^2}\hbar.
\end{equation} Together with equations (\ref{mq}) and (\ref{mp})
this leads to the expected uncertainty relations for $E$ and $T$.

So regardless of whether the box is moving freely or executes a
harmonic motion, and regardless of whether we base our predictions
on a determination of $q$ or on a determination of $p$, we always
find $\Delta E. \Delta T \geq \frac{1}{2}\hbar$. The uncertainties
in $q$ and $p$ on the one hand, and in $q_{cl}$ on the other,
guarantee that no conflict with the time-energy uncertainty
relation for the photon can arise. The essential point is that the
box quantities $q_{cl}$ and $m$ (as calculated from either $p$ or
$q$) form a complementary pair.

\section{Epilogue}

One might still wonder about the completely general case, with
arbitrary $V(q)$. The result of any specific calculation is known
beforehand, however: Any spread in the energy (or equivalently
mass) of the box, regardless of whether introduced by an
uncertainty $\Delta q$ or an uncertainty $\Delta p$, will result
in an uncertainty in the evolution of $q_{cl}$ by virtue of the
general time-energy relation \cite{messiah,busch,hilgevoord}
\begin{equation}\label{messiah,busch,hilgevoord}
\Delta H.\frac{\Delta q_{cl}}{\dot{<q_{cl}>}}\geq
\frac{1}{2}\hbar.
\end{equation} This guarantees in a general way
that the uncertainties will come out right. That does not make the
above specific calculations irrelevant, however. The latter show
in a bottom-up way how the danger of inconsistency is avoided by
the dynamics of quantum mechanics. The paradox that $p$ (or $q$)
and the clock time can be read off simultaneously, and that
therefore (seemingly) both $E_{ph}$ and $T_{arr}$ can be precisely
predicted, is dissolved by showing how the validity of
complementarity is guaranteed by the details of the dynamics.
These calculations put Bohr's historical, qualitative arguments
against Einstein on a firm quantitative basis.

Moreover, our calculations exhibit a ``fine structure'' in the
behavior of the uncertainties that does not follow from the
general uncertainty relation. As becomes clear from (\ref{unc1}),
(\ref{unc2}), (\ref{mq}), (\ref{mp}), (\ref{unc3}) and
(\ref{unc4}), the uncertainties in $E$ and $T$ are not time
independent. Although their products always satisfy the general
$E$-$T$ uncertainty relation, the distribution of uncertainty over
the quantities changes in time. This leads to a remarkable
conclusion: depending on \textit{when} the box measurement is
performed, the resulting quantum description of the photon
differs---in spite of the fact that in the meantime the photon can
have reached a distance of lightyears. This is another
illustration of the leading idea of the later EPR thought
experiment. We may safely assume that pondering the implications
of complementarity in the photon-box case has played a pivotal
role in Einstein's dissatisfaction with quantum mechanics. At the
same time, the experiment nicely illustrates how quantum mechanics
consistently deals with this type of correlated systems.

\end{document}